\documentclass[fleqn,twoside]{article}
\usepackage{espcrc2,amsmath}

\usepackage{graphicx}

\newcommand{\cO}{{\cal O}}

\newcommand{\as}{\alpha_s}
\newcommand{\wh}{\widehat}
\newcommand{\nn}{\nonumber}

\newcommand{\eqn}[1]{(\ref{#1})}

\newcommand{\gev}{\mbox{\rm GeV}}
\newcommand{\tvs}{\vbox{\vskip 4mm}}

\newcommand{\sfrac}[2]{\mbox{$\frac{#1}{#2}$}}
\title{Recent progress in hadronic $\tau$ decays}

\author{Matthias Jamin\address{
        Instituci\'o Catalana de Recerca i Estudis Avan\c{c}ats (ICREA),\\
        Institut de F\'isica d'Altes Energies (IFAE),
        Campus UAB, E-08193 Bellaterra, Barcelona, Spain}}

\begin{document}

\begin{abstract}
The determination of $\as$ from hadronic $\tau$ decays is impeded by the fact
that two choices for the renormalisation group resummation, namely fixed-order
(FOPT) and contour-improved perturbation theory (CIPT), yield systematically
differing results. On the basis of a model for higher-order terms in the
perturbative series, which incorporates well-known structure from renormalons,
it is found that FOPT smoothly approaches the Borel sum for the $\tau$ hadronic
width, while CIPT is unable to account for the resummed series. An example
for the behaviour of QCD spectral function moments, displaying a similar
behaviour, is presented as well. 
\vspace{1pc}
\end{abstract}

\maketitle

\section{INTRODUCTION}

Hadronic decays of the $\tau$ lepton provide an excellent playground for the
study of QCD at low energies. Its mass of $M_\tau\approx 1.8\,\gev$ is in an
energy region where perturbation theory is still applicable, but also
non-perturbative effects come into play and need to be included. These may
arise from vacuum-condensate terms in the framework of the operator product
expansion (OPE) and from so-called {\em duality-violating} contributions close
to the physical, Minkowskian energy axis.

In the seminal article \cite{bnp92}, the strategy for a precise determination
of the QCD coupling $\as$ from the total $\tau$ hadronic width
\begin{equation}
\label{Rtauex}
R_\tau \,\equiv\, \frac{\Gamma[\tau^- \to {\rm hadrons} \, \nu_\tau (\gamma)]}
{\Gamma[\tau^- \to e^- \overline \nu_e \nu_\tau (\gamma)]} \,=\,
3.640(10) \,,
\end{equation}
was developed, while in the subsequent years moments of spectral $\tau$-decay
distributions were incorporated into the analyses as well
\cite{aleph98,opal98,aleph05}.

The analytical computation of the perturbative order $\alpha_s^4$ correction
\cite{bck08} has recently revived the interest in $\alpha_s$ analyses from
hadronic $\tau$ decays which after evolution to the $Z$ boson mass scale
resulted in the following determinations:
\begin{equation}
\label{alphas}
\alpha_s(M_Z^2) =\begin{cases}
\;0.1202\, (6)_{\rm exp} (18)_{\rm
th}\,\,\,\,\,\mbox{\cite{bck08}}\,\ ,\cr
\;0.1212\, (5)_{\rm exp} (9)_{\rm th}\,\,\,\,\,\
\,\mbox{\cite{ddhmz08}}\,\
,\cr
\;0.1180\, (4)_{\rm exp} (7)_{\rm th}\,\,\,\,\, \
\,\mbox{\cite{bj08}}\
\,, \cr
\;0.1187\, (6)_{\rm exp} (15)_{\rm
th}\,\,\,\,\,\mbox{\cite{my08}}\
\,.
\end{cases}
\end{equation}
The dispersion in these results dominantly originates from different treatments
of the renormalisation group (RG) resummation of the perturbative series,
namely fixed-order perturbation theory (FOPT), or contour-improved perturbation
theory (CIPT) \cite{piv91,dp92}, being systematically larger than the last
included term in the expansion, which often in asymptotic series provides an
estimate of the uncertainty due to higher-order terms not included in the
partial sum. This points to the necessity of investigating the influence of
different RG resummations in more detail, which can for example be performed
within exactly resummable models for the perturbative series.

Most suitable for the $\as$ determination is the $\tau$ decay rate into light
$u$ and $d$ quarks $R_{\tau,V/A}$ via a vector or axialvector current, since
in this case power corrections are especially suppressed. Theoretically,
$R_{\tau,V/A}$ takes the form \cite{bnp92}
\begin{eqnarray}
\label{RtauVA}
R_{\tau,V/A} &\!\!=\!\!& \frac{N_c}{2}\,S_{\rm EW}\,|V_{ud}|^2\,\Big[\,
1 + \delta^{(0)} \nn \\
&& +\,\delta_{\rm EW}' + \sum\limits_{D\geq 2} \delta_{ud,V/A}^{(D)} \,\Big]\,,
\end{eqnarray}
where $S_{\rm EW}=1.0198(6)$ \cite{ms88} and $\delta_{\rm EW}'=0.0010(10)$
\cite{bl90} are electroweak corrections, $\delta^{(0)}$ comprises the
perturbative QCD correction, and the $\delta_{ud,V/A}^{(D)}$ denote quark
mass and higher $D$-dimensional operator corrections which arise in the
framework of the OPE.

\begin{boldmath}
\section{PERTURBATIVE CORRECTION $\delta^{(0)}$}
\end{boldmath}

Below, only the purely perturbative correction $\delta^{(0)}$ shall be
considered, which gives the dominant contribution to $R_{\tau,V/A}$. In FOPT
it takes the general form
\begin{equation}
\label{del0FO}
\delta^{(0)}_{\rm FO} \,=\, \sum\limits_{n=1}^\infty a(M_\tau^2)^n
\sum\limits_{k=1}^{n} k\,c_{n,k}\,J_{k-1} \,,
\end{equation}
where $a(\mu^2)\equiv a_\mu\equiv\as(\mu)/\pi$, and $c_{n,k}$ are the
coefficients which appear in the perturbative expansion of the vector
correlation function,
\begin{equation}
\label{Pis}
\Pi_V(s) \,=\, -\,\frac{N_c}{12\pi^2} \sum\limits_{n=0}^\infty a_\mu^n
\sum\limits_{k=0}^{n+1} c_{n,k} \ln^k\!\left(\frac{-s}{\mu^2}\right) .
\end{equation}
At each perturbative order, the coefficients $c_{n,1}$ can be considered
independent, while all other $c_{n,k}$ with $k\geq 2$ are calculable from the
RG equation. Further details can for example be found in ref.~\cite{bj08}.
Finally, the $J_l$ are contour integrals in the complex $s$-plane, which are
defined by
\begin{equation}
\label{Jl}
J_l \,\equiv\, \frac{1}{2\pi i} \!\!\oint\limits_{|x|=1} \!\!
\frac{dx}{x}\, (1-x)^3\,(1+x) \ln^l(-x) \,.
\end{equation}
The first three, being required up to $\cO(\as^3)$, take the numerical
values
\begin{equation}
\label{J0to2}
J_0 \,=\, 1 \,, \quad
J_1 \,=\, -\,\sfrac{19}{12} \,, \quad
J_2 \,=\, \sfrac{265}{72} - \sfrac{1}{3}\,\pi^2 \,.
\end{equation}

At order $\as^n$ FOPT contains unsummed logarithms of order 
$\ln^l(-x)\sim \pi^l$ with $l<n$ related to the contour integrals $J_l$.
CIPT sums these logarithms, which yields
\begin{equation}
\label{del0CI}
\delta^{(0)}_{\rm CI} \,=\, \sum\limits_{n=1}^\infty c_{n,1}\,
J_n^a(M_\tau^2)
\end{equation}
in terms of the contour integrals $J_n^a(M_\tau^2)$ over the running coupling,
defined as:
\begin{equation}
\label{Jna}
J_n^a(M_\tau^2) \,\equiv\, \frac{1}{2\pi i} \!\!\oint\limits_{|x|=1}\!\!
\frac{dx}{x}\,(1-x)^3\,(1+x)\,a^n(-M_\tau^2 x) \,.
\end{equation}
In contrast to FOPT, for CIPT each order $n$ just depends on the corresponding
coefficient $c_{n,1}$. Thus, all contributions proportional to the coefficient
$c_{n,1}$ which in FOPT appear at all perturbative orders equal or greater than
$n$ are resummed into a single term.

Numerically, the two approaches lead to significant differences. Employing the
recent average $\as(M_Z)= 0.1184$ \cite{bet09}, leading to $\as(M_\tau)=0.3186$,
in eqs.~\eqn{del0FO} and \eqn{del0CI}, one finds
\begin{eqnarray}
\label{del0FOn}
\delta^{(0)}_{\rm FO} &\!\!=\!\!& 0.1959 \;(0.2022) \,, \\
\tvs
\label{del0CIn}
\delta^{(0)}_{\rm CI} &\!\!=\!\!& 0.1814 \;(0.1847) \,,
\end{eqnarray}
where the first number in both cases employs the known coefficients up to
$\cO(\as^4)$ \cite{bck08} and the numbers in brackets include an estimate of
the $\cO(\as^5)$ term with $c_{5,1}\approx 283$ \cite{bj08}. Inspecting the
individual contributions from each order, up to $\cO(\as^5)$ the CIPT series
appears to be better convergent. However, around the seventh order, the
contour integrals $J_n^a(M_\tau^2)$ change sign and thus at this order the
contributions are bound to become small. Therefore, the faster approach to the
minimal term does not necessarily imply that CIPT gives the closer approach to
the true result for the resummed series.

\vskip 6mm 
\section{A PHYSICAL MODEL}

To investigate whether FOPT or CIPT results in a better approximation to
$\delta^{(0)}$, one requires a physically motivated model for its series.
Such a model was constructed in ref.~\cite{bj08} and is based on the Borel
transform of the Adler function $D_V(s)$:
\begin{equation}
\label{DVs}
D_V(s) \,\equiv\, -\,s\,\frac{d}{ds}\,\Pi_V(s) \,\equiv\,
\frac{N_c}{12\pi^2}\,\big[ 1+\wh D(s) \big] \,.
\end{equation}
In the following discussion it is slightly more convenient to utilise the
related function $\wh D(s)$. Its Borel transform $B[\wh D](t)$ is defined by
the relation
\begin{equation}
\label{Dalpha}
\wh D(\alpha) \,\equiv\, \int\limits_0^\infty dt\,{\rm e}^{-t/\alpha}\,
B[\wh D](t)\,.
\end{equation}
The integral $\wh D(\alpha)$, if it exists, gives the Borel sum of the original
divergent series. It was found that the Borel-transformed Adler function
$B[\wh D](t)$ obtains infrared (IR) and ultraviolet (UV) renormalon poles at
positive and negative integer values of the variable $u\equiv 9t/(4\pi)$,
respectively \cite{ben93,bro93,ben98}. (With the exception of $u=1$.)

Apart from very low orders, where a dominance of renormalon poles close to
$u=0$ has not yet set in, intermediate orders should be dominated by the
leading IR renormalon poles, while the leading UV renormalon, being closest
to $u=0$, dictates the large-order behaviour of the perturbative expansion.
Assuming that only the first two orders are not yet dominated by the lowest
IR renormalons, one is led to the ansatz
\begin{eqnarray}
\label{BRu}
\hspace{-2mm} B[\wh D](u) &\!\!=\!\!& B[\wh D_1^{\rm UV}](u) +
B[\wh D_2^{\rm IR}](u) + \nn \\
\tvs
&& B[\wh D_3^{\rm IR}](u) + d_0^{\rm PO} + d_1^{\rm PO} u \,,
\end{eqnarray}
which includes one UV renormalon at $u=-1$, the two leading IR renormalons at
$u=2$ and $u=3$, as well as polynomial terms for the two lowest perturbative
orders. Explicit expressions for the UV and IR renormalon pole terms
$B[\wh D_p^{\rm UV}](u)$ and $B[\wh D_p^{\rm IR}](u)$ can be found in
section~5 of ref.~\cite{bj08}.

Apart from the residues $d_p^{\rm UV}$ and $d_p^{\rm IR}$, the full structure
of the renormalon pole terms is dictated by the OPE and the RG. Therefore, the
model~\eqn{BRu} depends on five parameters, the three residua $d_1^{\rm UV}$,
$d_2^{\rm IR}$ and $d_3^{\rm IR}$, as well as the two polynomial parameters
$d_0^{\rm PO}$ and $d_1^{\rm PO}$. These parameters can be fixed by matching
to the perturbative expansion of $\wh D(s)$ up to $\cO(\as^5)$. Thereby, also
the estimate for $c_{5,1}$ is used. The parameters of the model \eqn{BRu} are
then found to be:
\begin{equation}
\label{dUVIRa}
d_1^{\rm UV} =\, -\,1.56\cdot 10^{-2} \,,\;
d_2^{\rm IR} =\,    3.16  \,,\;
d_3^{\rm IR} =\, -\,13.5 \,, \nn \\[-1mm]
\end{equation}
\begin{equation}
\label{dUVIRb}
d_0^{\rm PO} =\,    0.781 \,,\;
d_1^{\rm PO} =\,    7.66\cdot 10^{-3} \,.
\end{equation}
The fact that the parameter $d_1^{\rm PO}$ turns out to be small implies that
the coefficient $c_{2,1}$ is already reasonably well described by the
renormalon pole contribution, although it was not used to fix the residua.
Therefore, one could set $d_1^{\rm PO}=0$ and actually work with a model which
only has four parameters. The predicted value $c_{5,1}=280$ then turns out
very close to the estimate, which can be viewed as one test of the stability
of the model.

\begin{table}[htb]
\renewcommand{\tabcolsep}{1.6mm} % set column spacing
\renewcommand{\arraystretch}{1.2} % enlarge line spacing
\begin{center}
\begin{tabular}{rrrrrr}
\hline
 & $c_{2,1}$ & $c_{3,1}$ & $c_{4,1}$ & $c_{5,1}$ & $c_{6,1}$ \\
\hline
${\rm IR}_2$ & $-77.8$ & $ 82.4$ & $100.4$ & $135.9$ & $ 97.5$ \\
${\rm IR}_3$ & $152.0$ & $ 28.7$ & $-10.0$ & $-20.2$ & $-13.3$ \\
${\rm UV}_1$ & $ 22.5$ & $-11.2$ & $  9.7$ & $-15.6$ & $ 15.8$ \\
\hline
\end{tabular}\\[2mm]
\end{center}
\caption{Relative contributions (in \%) of the different IR and UV renormalon
poles to the Adler-function coefficients $c_{2,1}$ to $c_{6,1}$ for the
Borel model \eqn{BRu}.\label{tab1}}
\vspace{-6mm}
\end{table}

This is also corroborated in table~1, where the relative contributions of a
certain renormalon pole to the coefficients $c_{2,1}$ to $c_{6,1}$ is tabulated.
The sum of the contributions to $c_{2,1}$ is close to 100\%, implying again
that the polynomial term is already small. Then, from $c_{3,1}$ to $c_{6,1}$
the leading IR pole at $u=2$ is dominating the coefficients, before the
leading UV pole at $u=-1$ takes over at even higher orders. (For the central
model \eqn{BRu} this happens around the 10th order \cite{bj08}.)

%%%%%%
\begin{figure}[thb]
\vspace{-2mm} \hspace{-3mm}
\includegraphics[angle=0, width=7.8cm]{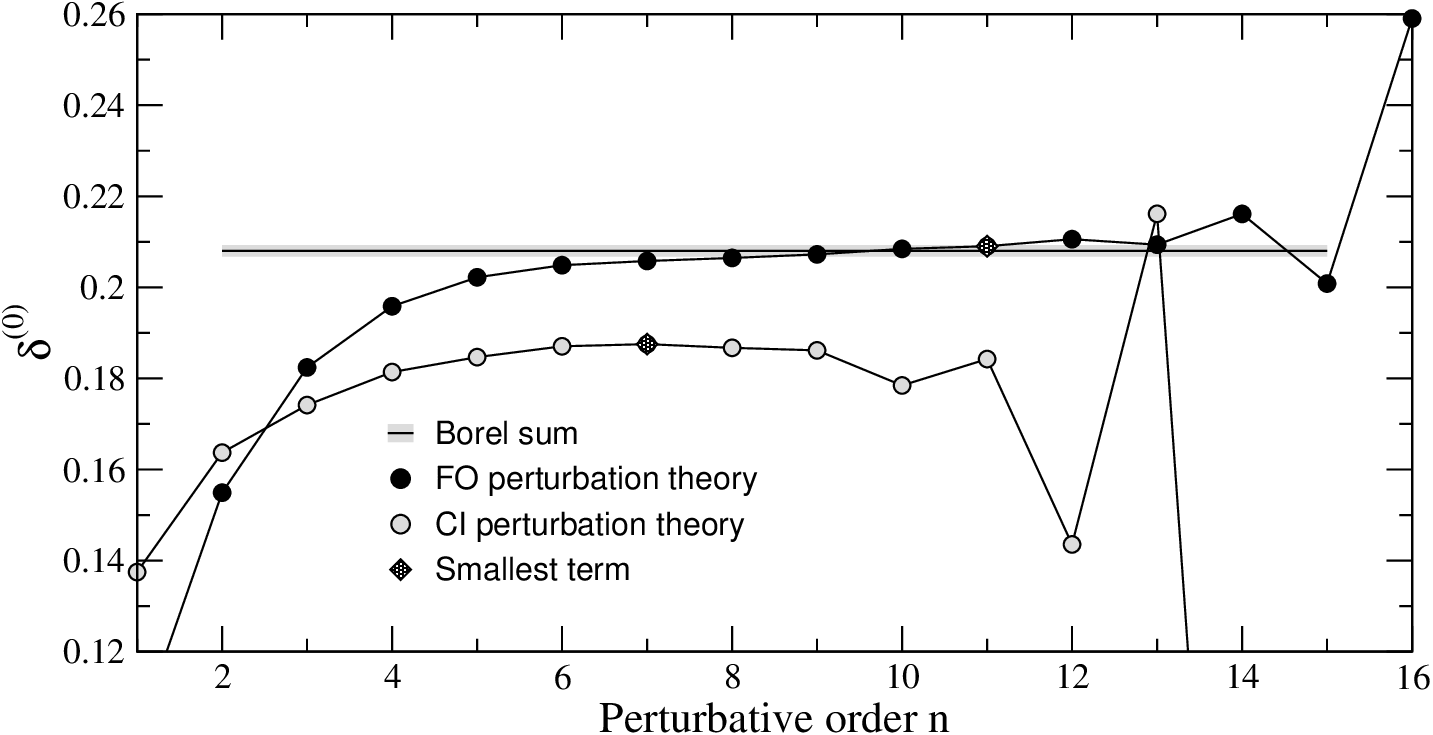}
\vspace{-1cm}
\caption{Results for $\delta^{(0)}_{\rm FO}$ (full circles) and
$\delta^{(0)}_{\rm CI}$ (grey circles) at $\as(M_\tau)=0.3186$, employing the
model \eqn{BRu}, as a function of the order $n$ up to which the terms in the
perturbative series have been summed. The straight line represents the result
for the Borel sum of the series.\label{fig1}}
\vspace{-4mm}
\end{figure}
%%%%%%

The implications of the model \eqn{BRu} for $\delta^{(0)}$ in FOPT and CIPT is
graphically represented in figure~\ref{fig1}. The full circles denote the
result for $\delta^{(0)}_{\rm FO}$ and the grey circles the one for
$\delta^{(0)}_{\rm CI}$, as a function of the order $n$ up to which the
perturbative series has been summed. The straight line corresponds to the
principal value Borel sum of the series, $\delta^{(0)}_{\rm BS}=0.2080$, and
the shaded band provides an error estimate based on its imaginary part divided
by $\pi$. The order at which the series have their smallest terms is indicated
by the grey diamonds. As is obvious from figure~\ref{fig1}, FOPT displays the
behaviour expected from an asymptotic series: the terms decrease up to a
certain order around which the closest approach to the resummed result is
found, and for even higher orders, the divergent large-order behaviour of the
series sets in. For CIPT, on the other hand, the asymptotic behaviour sets in
earlier, and the series is never able to come close to the Borel
sum.\footnote{The same conclusions had already been drawn in ref.~\cite{bbb95}
on the basis of the large-$\beta_0$ approximation for $\Pi_V(s)$.}

\begin{table}[htb]
\renewcommand{\tabcolsep}{1.6mm} % set column spacing
\renewcommand{\arraystretch}{1.2} % enlarge line spacing
\begin{center}
\begin{tabular}{rrrrrr}
\hline
 & $c_{2,1}$ & $c_{3,1}$ & $c_{4,1}$ & $c_{5,1}$ & $c_{6,1}$ \\
\hline
${\rm IR}_3$ & $-743.3$ & $-140.5$ & $49.1$ & $98.9$ & $99.1$ \\
${\rm IR}_4$ & $ 662.8$ & $ 244.2$ & $47.7$ & $ 6.3$ & $-7.2$ \\
${\rm UV}_1$ & $   7.5$ & $  -3.7$ & $ 3.2$ & $-5.2$ & $ 8.1$ \\
\hline
\end{tabular}\\[2mm]
\end{center}
\caption{Relative contributions (in \%) of the different IR and UV renormalon
poles to the Adler-function coefficients $c_{3,1}$ to $c_{6,1}$ for the Borel
model with $d_2^{\rm IR}=0$.\label{tab2}}
\vspace{-6mm}
\end{table}

As the behaviour of CIPT versus FOPT hinges on the contribution of the leading
IR renormalon at $u=2$, in principal also models can be constructed for which
CIPT provides a better account of the Borel sum. These would generally be
models where $d_2^{\rm IR}$ is much smaller than the value quoted in
eq.~\eqn{dUVIRb}. While such models can at present not be excluded, the pattern
of the individual contributions appears more unnatural than in the main model
\eqn{BRu}: the known $c_{n,1}$ can only be reproduced when one allows for large
cancellations between the individual terms. Thus, the behaviour generally
expected from the presence of renormalon poles, namely dominance of leading
IR poles at intermediate orders, would be lost.

This is apparent from table~\ref{tab2}, which is the analog of table~\ref{tab1},
but for a model where $d_2^{\rm IR}$ is forced to be zero and an additional
IR pole at $u=4$ is added, in order to be able to reproduce the known Adler
function coefficients. On the one hand for low coefficients there are huge
cancellations between the IR renormalon poles and also $c_{2,1}$ is not well
described at all. This entails that a large, additional polynomial term is
required.  Besides, it appears unnatural that the residue of the first IR
renormalon pole at $u=2$ is small, and still there should be a natural size
contribution of the gluon condensate to the Adler function.

\begin{figure}[thb]
\vspace{-2mm} \hspace{-3mm}
\includegraphics[angle=0, width=7.8cm]{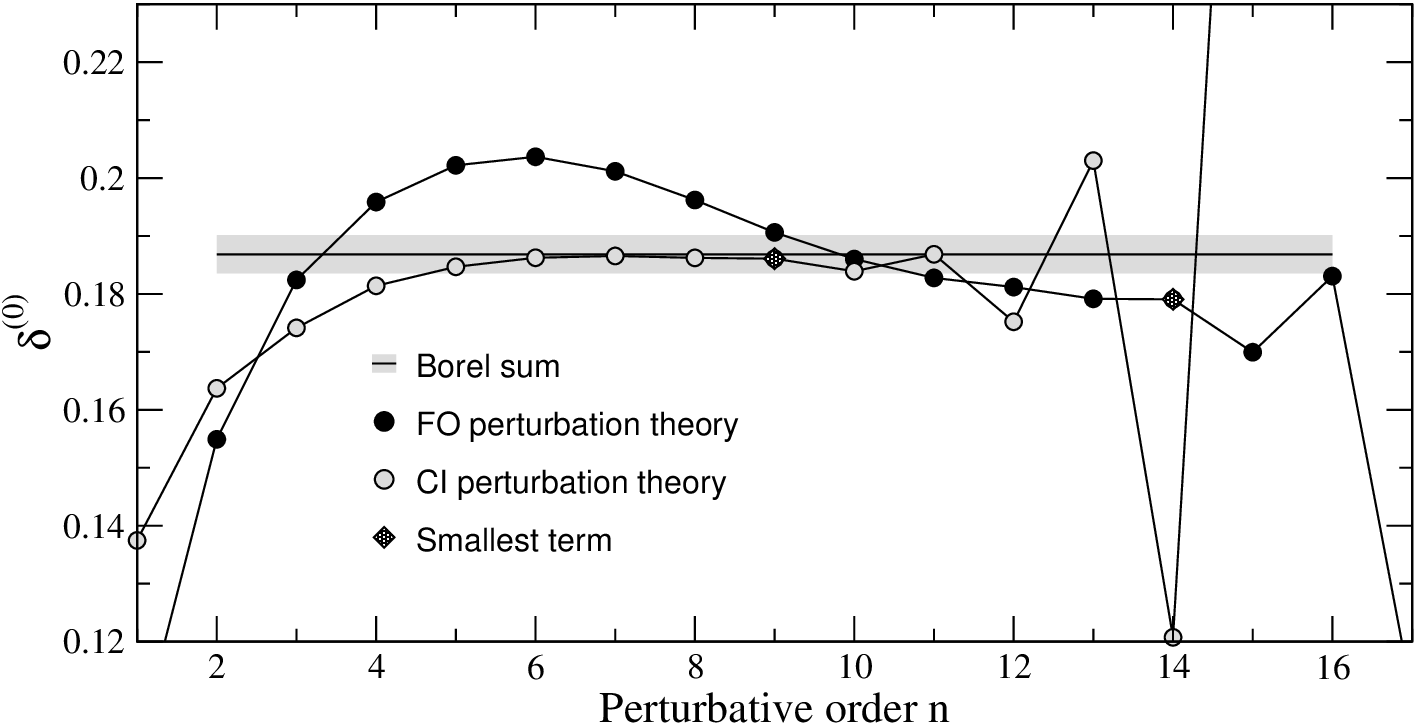}
\vspace{-1cm}
\caption{Results for $\delta^{(0)}_{\rm FO}$ (full circles) and
$\delta^{(0)}_{\rm CI}$ (grey circles) at $\as(M_\tau)=0.3186$, employing the
model \eqn{BRu} with $d_2^{\rm IR}=0$ and an additional IR pole at $u=4$, as
a function of the order $n$ up to which the terms in the perturbative series
have been summed. The straight line represents the result for the Borel sum
of the series.\label{fig2}}
\vspace{-4mm}
\end{figure}

A graphical account of the model with $d_2^{\rm IR}=0$ is presented in
figure~\ref{fig2}. As anticipated, now CIPT provides a good description of
the Borel sum, while FOPT is able to come reasonably close to it around its
minimal term, but generally is rather badly behaved.\footnote{A similar
behaviour was found in ref.~\cite{jam05} in models where the higher-order
Adler function coefficients were assumed to be small, in contrast to the
expected asymptotic behaviour of this series in QCD.} Nonetheless, again, the
behaviour observed in table~\ref{tab2} and figure~\ref{fig2} appears unnatural
from the perspective of the structure of the Borel transform of the Adler
function and should be considered less likely than the behaviour of the
model \eqn{BRu} with the residues \eqn{dUVIRb}.

%%%%%%
%%%%%%

In the standard $\as$ determinations from hadronic $\tau$ decays
\cite{aleph98,opal98}, besides the total decay rate also moments of the
spectral decay distributions are employed. Historically, the so-called
$(k,l)$-moments were used, for which a polynomial $(1-x)^k x^l$ is multiplied
to the kinematical weight function from phase-space. In the analyses
\cite{aleph98,opal98}, moments with $k=1$ and $l=0,1,2,3$ were taken into
account, such that also condensate contributions up to dimension-8 could be
extracted in addition to $\as$. Therefore, it is of interest to analyse the
behaviour of these moments in models of higher orders of perturbation theory
as well.

%%%%%%
\begin{figure}[thb]
\vspace{-2mm} \hspace{-3mm}
\includegraphics[angle=0, width=7.8cm]{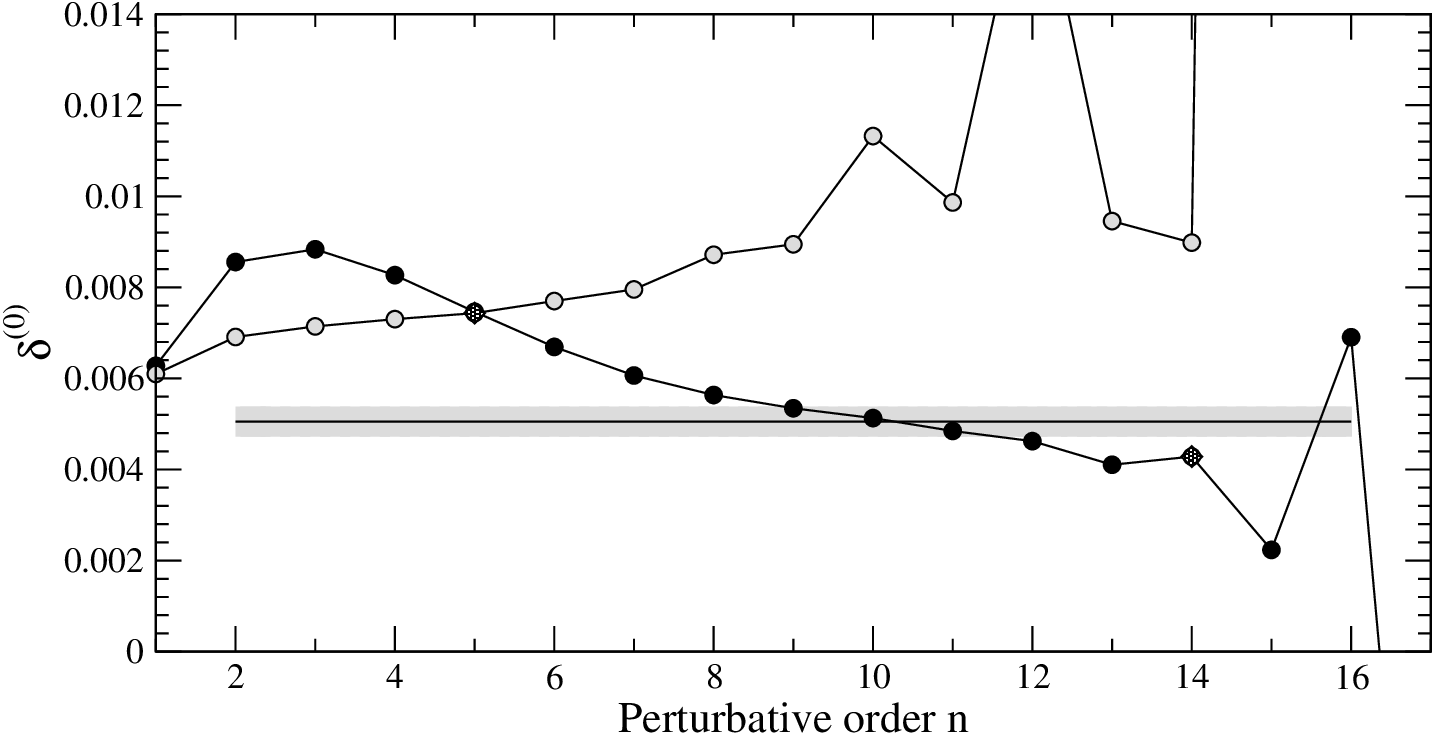}
\vspace{-1cm}
\caption{Results for $(1,2)$ moments $\delta^{(0,12)}_{\rm FO}$
(full circles) and $\delta^{(0,12)}_{\rm CI}$ (grey circles) at
$\as(M_\tau)=0.3186$, employing the model \eqn{BRu}, as a function of the
order $n$ up to which the terms in the perturbative series have been summed.
The straight line represents the result for the Borel sum of the series.
\label{fig3}}
\vspace{-4mm}
\end{figure}
%%%%%%

An example of such an analysis is shown in figure~\ref{fig3} for the moment
$(1,2)$. It is observed that, like for $\delta^{(0)}$, CIPT is unable to
provide a reasonable account of the full Borel sum. In contrast, FOPT comes
close to the resummed result for perturbative orders around the minimal term,
though it is obvious that this particular moment displays a very bad behaviour
of the asymptotic series, with only unsatisfactory convergence up to its
minimal term. The example of the $(1,2)$-moment should motivate an exhaustive
investigation of the moments employed in previous $\as$ analyses from hadronic
$\tau$ decays, which will be presented in the near future.

\section{CONCLUSIONS}

Models of higher orders of perturbation theory allow for the study of different
resummation prescriptions in the computation of the total $\tau$ hadronic width,
as well as related moments of $\tau$ decay spectral distributions. The most
prominent methods are fixed-order perturbation theory (FOPT) and the
so-called contour-improved perturbation theory (CIPT), which performs a partial
resummation of running effects of the QCD coupling $\as$ in the integration
along the complex contour in the $s$-plane.

A physically motivated model for the higher-order behaviour of the Adler
function was presented in ref.~\cite{bj08}, and is given in eq.~\eqn{BRu}.
The model was based on the general structure of the Borel transform of the
Adler function and the renormalisation group equation. Furthermore, knowledge
on the first four analytically available coefficients $c_{1,1}$ to $c_{4,1}$,
as well as an estimate for the fifth coefficient $c_{5,1}$, were incorporated.

Results for $\delta^{(0)}$ in the main model \eqn{BRu} were displayed in
figure~\ref{fig1}, and it is observed that while FOPT provides a good account
of the full Borel summation, CIPT is never able to come close to the resummed
value.\footnote{The general behaviour of the Borel model is also supported by an
independent approach where the perturbative series is conformally transformed
into a series which displays better convergence properties than the original
series in powers of $\as$ \cite{cf09}, though in this case a modified CIPT
better converges towards the full result.} This general behaviour hinges on
the size of the residue of the first IR renormalon pole at $u=2$. In a model
in which this residue is set to zero by hand, on the contrary CIPT well
describes the Borel sum, whereas FOPT, though approaching the resummed value
around its minimal term, generally is rather badly behaved. Models in which
$d_2^{\rm IR}\approx 0$, however, are only able to reproduce the known Adler
function coefficients through large cancellations between different IR
renormalon contributions, which appears unnatural.

In the standard experimental extractions of $\as$ from hadronic $\tau$
decays, also moments of the decay spectra are employed \cite{aleph98,opal98}.
An example of the behaviour of such a moment was shown in figure~\ref{fig3}.
The repeatedly unacceptable behaviour of CIPT in this case and the also
unsatisfactory convergence of FOPT, suggest that such moments should be
investigated systematically, before their usefulness in $\as$ determinations
from hadronic $\tau$ decays is corroborated.

A final topic, not touched upon at all so far, are violations of quark-hadron
duality \cite{bsz01,cgp05,cgp08}. Like the perturbative expansion in $\as$,
also the OPE in inverse powers of $s$ may be only asymptotic, and thus
exponentially suppressed terms could be relevant close to the Minkowskian axis
where the bound states are situated. This was investigated in a model in
\cite{cgp05} and the possible influence of duality violations in $\tau$ decay
spectra was studied in refs.~\cite{cgp08}. Also here a more systematic
investigation seems to be in order which was initiated in \cite{bcgjmop10} and
will be continued in the future.

Hadronic $\tau$ decays have proven to be a very fruitful laboratory for the
study of low-energy QCD and the extractions of fundamental QCD parameters
like the coupling $\as$. However, there remain unresolved theoretical issues
which taint the precision of these determinations. Numerically the two most
relevant appear to be the resummation of QCD running effects in the computation
of the $\tau$ hadronic width and related decay moments as well as duality
violations. As far as the former topic is concerned, a physically motivated
model of the QCD Adler function favours the use of FOPT, since generally it
is better able to represent the fully resummed series.

\vskip 4mm 
\section*{Acknowledgements}
The author would like to thank Martin~Beneke for a most enjoyable collaboration.
This work has been supported in parts by the Spanish Ministry (grants
CICYT-FEDER FPA2007-60323, FPA2008-01430, CPAN CSD2007-00042), by the Catalan
Government (grant SGR2009-00894), and by EU Contract MRTN-CT-2006-035482
(FLAVIAnet).

\providecommand{\href}[2]{#2}\begingroup\raggedright\endgroup

\vfill

\end{document}